\documentclass[conference]{IEEEtran}
\usepackage{cite}
\usepackage{amsmath,amssymb,amsfonts}
\usepackage{algorithmic}
\usepackage{graphicx}
\usepackage{textcomp}
\usepackage{hyperref}
\usepackage{etoolbox}
\usepackage{ifthen}

\usepackage[letterpaper, left=0.75in, right=0.75in, top=0.75in, bottom=1.1in]{geometry}

\def\BibTeX{{\rm B\kern-.05em{\sc i\kern-.025em b}\kern-.08em
    T\kern-.1667em\lower.7ex\hbox{E}\kern-.125emX}}

\newboolean{proceedings}
\setboolean{proceedings}{false}  

\newcommand{\fig}[1]{Fig.\,\ref{fig:#1}}
\newcommand{\eq}[1]{Eq.\,\eqref{eq:#1}}
\newcommand{\secref}[1]{Sec.\,\ref{sec:#1}}

\ifthenelse{\boolean{proceedings}}{
  \newcommand{\appRef}[1]{Appendix~\ref{app:#1} in \cite{arxiv}}
}{
  \newcommand{\appRef}[1]{Appendix~\ref{app:#1}}
}

\newcommand{\authorswitch}[1]{%
  \ifcase#1
\author{
\IEEEauthorblockN{Chetan Gohil$^{\dagger}$}
\IEEEauthorblockA{\textit{Centre for Complex Systems} \\
\textit{School of Computer Science}\\
\textit{Brain and Mind Centre}\\
\textit{The University of Sydney}\\
Sydney, New South Wales, Australia \\
chetan.gohil@sydney.edu.au} \\
\and
\IEEEauthorblockN{Oliver M. Cliff$^{\dagger}$}
\IEEEauthorblockA{\textit{Centre for Complex Systems} \\
\textit{School of Computer Science} \\
\textit{School of Physics} \\
\textit{The University of Sydney}\\
Sydney, New South Wales, Australia \\
oliver.cliff@sydney.edu.au}
\and
\IEEEauthorblockN{James M. Shine}
\IEEEauthorblockA{\textit{Brain and Mind Centre} \\
\textit{The University of Sydney}\\
Sydney, New South Wales, Australia \\
mac.shine@sydney.edu.au}
\and
\IEEEauthorblockN{Ben D. Fulcher}
\IEEEauthorblockA{\textit{Centre for Complex Systems} \\
\textit{School of Physics} \\
\textit{The University of Sydney}\\
Sydney, New South Wales, Australia \\
ben.fulcher@sydney.edu.au}
\and
\IEEEauthorblockN{Joseph T. Lizier}
\IEEEauthorblockA{\textit{Centre for Complex Systems} \\
\textit{School of Computer Science} \\
\textit{The University of Sydney}\\
Sydney, New South Wales, Australia \\
joseph.lizier@sydney.edu.au}
}   
  \or
\author{
\IEEEauthorblockN{Chetan Gohil$^{~||}$,
\IEEEauthorrefmark{1},
\IEEEauthorrefmark{2},
\IEEEauthorrefmark{4},
Oliver M. Cliff$^{~||}$,
\IEEEauthorrefmark{1},
\IEEEauthorrefmark{3},
\IEEEauthorrefmark{4},
James M. Shine
\IEEEauthorrefmark{1},
\IEEEauthorrefmark{2},
Ben D. Fulcher
\IEEEauthorrefmark{1},
\IEEEauthorrefmark{3},
and Joseph T. Lizier
\IEEEauthorrefmark{1},
\IEEEauthorrefmark{4}}
\IEEEauthorblockA{\IEEEauthorrefmark{1}School of Computer Science,}
\IEEEauthorblockA{\IEEEauthorrefmark{2}Brain and Mind Centre,}
\IEEEauthorblockA{\IEEEauthorrefmark{3}School of Physics, and\\
\IEEEauthorblockA{\IEEEauthorrefmark{4}Centre for Complex Systems}
The University of Sydney\\
Sydney, New South Wales, Australia\\
Email: chetan.gohil@sydney.edu.au}
}
  \or
\author{}
  \fi
}

\begin{document}

\title{Cross Mutual Information}

\authorswitch{1}  

\maketitle
\ifthenelse{\boolean{proceedings}}{
} {
  \thispagestyle{plain}
  \pagestyle{plain}
}

\renewcommand{\thefootnote}{\fnsymbol{footnote}}
\footnotetext{$^{||}$Equal contribution.}

\begin{abstract}
Mutual information (MI) is a useful information-theoretic measure to quantify the statistical dependence between two random variables: $X$ and $Y$. Often, we are interested in understanding how the dependence between $X$ and $Y$ in one set of samples compares to another. Although the dependence between $X$ and $Y$ in each set of samples can be measured separately using MI, these estimates cannot be compared directly if they are based on samples from a non-stationary distribution. Here, we propose an alternative measure for characterising how the dependence between $X$ and $Y$ as defined by one set of samples is expressed in another, \textit{cross mutual information}. We present a comprehensive set of simulation studies sampling data with $X$-$Y$ dependencies to explore this measure. Finally, we discuss how this relates to measures of model fit in linear regression, and some future applications in neuroimaging data analysis.
\end{abstract}


\section{Introduction} \label{sec:intro}
Mutual information (MI)~\cite{thomas2006} is a useful information-theoretic measure for the dependence between two random variables: $X$ and $Y$. It quantifies the amount of information that observing one variable provides about the other. For continuous-valued data, MI is defined as:
\begin{equation} \label{eq:mi-continuous}
I(X; Y) = \int_{y \in Y} \int_{x \in X} p(x,y) \, \log \left( \frac{p(x,y)}{p(x)p(y)} \right) \mathrm{d}x \, \mathrm{d}y,
\end{equation}
where $p(x,y)$ is the joint probability distribution for the random variables, $p(x) = \int p(x,y) \, \mathrm{d}y$, $p(y) = \int p(x,y) \, \mathrm{d}x$ are their marginal probability distributions, and $x$ ($y$) is a sample of the variable $X$ ($Y$). The joint probability distribution $p(x,y)$ defines a data generating system we can observe, i.e. sample. MI can be estimated based on a set of samples using an estimator, such as KSG~\cite{kraskov2004}, to understand the dependence between $X$ and $Y$ in the system.

We can write the MI in \eq{mi-continuous} (or similarly for discrete-valued variables) as an expectation over the sample space,
\begin{equation} \label{eq:mi}
I_p(X;Y) = \mathbb{E}_{p} \{ i_p(x;y) \}.
\end{equation} 
where
\begin{equation} \label{eq:local-mi}
i_p(x;y) = \log \left( \frac{p(x,y)}{p(x)p(y)} \right)
\end{equation}
is the \textit{local} (or \textit{pointwise}) MI \cite{fano1961}. Here we have used the notation $I_p$ and $i_p$ to highlight that this is a function of the underlying probability distribution $p$ in the subscript.

\begin{figure*}
    \centering
    \includegraphics[width=0.91\textwidth]{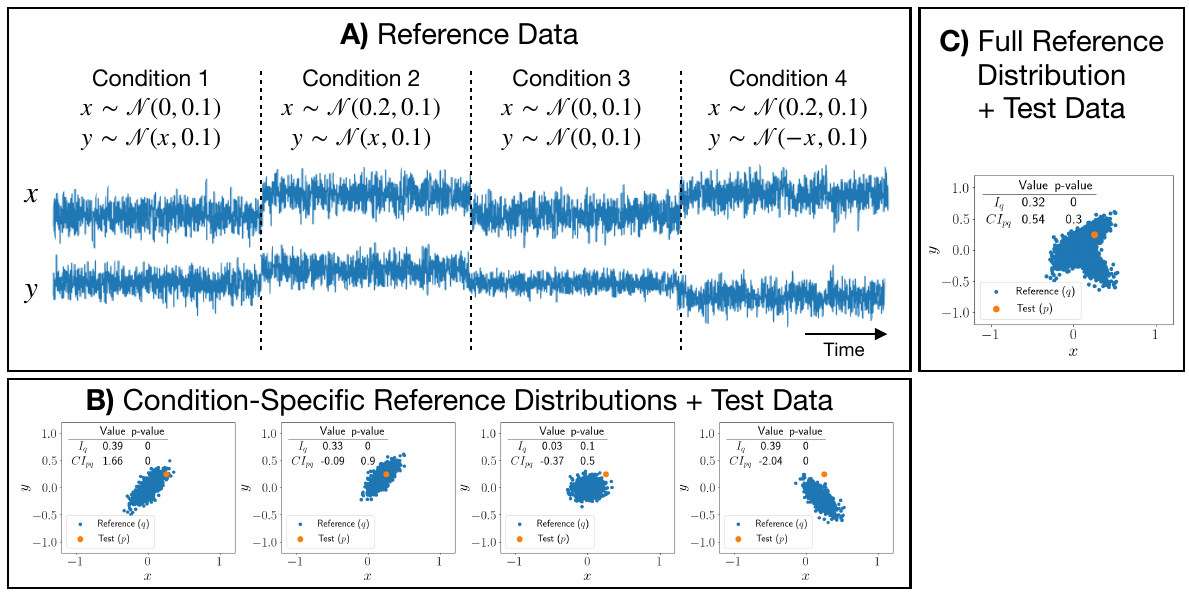}
    \caption{\textbf{The cross MI allows us to handle non-stationary and online data}. A) Simulation of a reference system that switches between different conditions (400 samples per condition). Each condition corresponds to a different dependency between the channels $X$ and $Y$. B) Scatter plots of the data from each condition (reference data, blue) and a new `online' (test) data point (orange, X=0.25, Y=0.25 in all cases) with its cross MI. C) Scatter plot of all the reference data (blue) and new test data point (orange) with its cross MI.}
    \label{fig:state-switching}
\end{figure*}

In some scenarios the system we observe is in a \textit{conditioned} state, where data is generated from a particular region of the sample space. We illustrate this in \fig{state-switching}, where the system transitions between different conditions with different $X$-$Y$ dependencies; such a system is referred to as being \textit{non-stationary}. Often, we assume a system is \textit{ergodic}, i.e., with time the system explores all possible conditions~\cite{thomas2006}.

The question we consider in this work is how do we compare the dependence between $X$ and $Y$ in one condition (or more generally, set of samples) to another condition (or set of samples). Naively, we can compare the MI estimated using the data from each condition in isolation. However, these estimates would be conditional MIs, which can remove redundant information or incorporate synergistic information with each condition in misleading ways -- discussed further in \secref{mi-issues}. This approach also does not account for the relative likelihoods of the different conditions (and dependencies between $X$ and $Y$) for the system to exhibit. In some cases, we may only have one sample from the condition of interest, e.g., with `online' systems that collect data on a live ongoing basis. Here, it is not be possible to calculate the conditional MI for the new data because the underlying probability distribution cannot be estimated accurately with a single data point.

To provide an alternative to address the issues related to conditioning, we propose a new information-theoretic measure, which we call the \textit{cross mutual information} (cross MI) in analogy to the well-known \textit{cross entropy}~\cite{murphy2022}. This quantity allows us to measure how strongly an $X$-$Y$ dependence defined by a \textit{reference} distribution ($q$) is expressed in new \textit{test} data (sampled from $p$).

The cross MI addresses some key challenges in estimating the relationship between two variables in non-stationary or online systems. Measuring the conventional MI $I_p$ for $p$ requires us to have sufficient data to estimate the test probability distribution $p$. In scenarios where we have a limited number of test samples or when $p$ is non-stationary this is not the case. This occurs in sliding window analyses, where we want to estimate the relationship between two variables using a small number of test samples in order to be sensitive to changes in the relationship, but at the same time use a well-sampled probability distribution for robustness.  Alternatively, in online applications, we may not want to or be able to update $p$ as new data comes in. In such cases, an alternative (perhaps well-sampled and/or offline) reference distribution $q$ can be leveraged using the cross MI. Moreover, in systems that dynamically switch between conditions this allows us to measure the strength of a relationship in a specific system condition as $p$ relative to the ensemble distribution over all conditions as $q$.

We introduce the cross MI in \secref{theory} and present a comprehensive set of simulations exploring how this measure behaves in a variety of regimes in \secref{simulations}. Finally, we discuss future applications and some important considerations when applying this measure in \secref{conclusions}.

\section{Theory} \label{sec:theory}
Here, we highlight some shortcomings of a conventional MI analysis when comparing different system conditions and present the cross MI. Statistical significance testing for checking whether a particular MI or cross MI value can be obtained by chance when there is no dependence is discussed in \appRef{statSigTesting}.

\subsection{Issues Comparing MI} \label{sec:mi-issues}
\begin{figure*}
    \centering
    \includegraphics[width=0.87\textwidth]{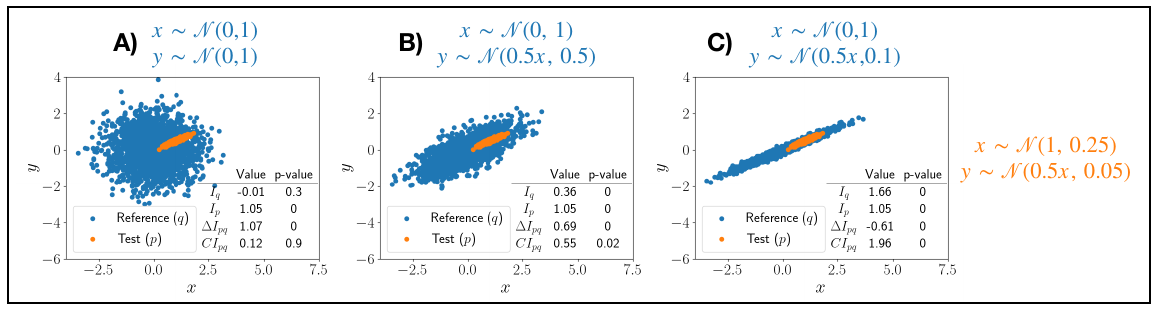}
    \caption{\textbf{Behaviour of the cross MI for different reference distributions that support the test data}. Difference choices for the reference data (blue, 2000 samples) are shown in A, B and C. All choices span the range of the test data (orange, 500 samples). Note, the negative value for $I_q$ in (A) is due to estimation noise.}
    \label{fig:inside-support}
\end{figure*}

A conventional approach in determining whether the $X$-$Y$ dependence in two system conditions is the same or different is to directly compare MI estimates obtained in each condition. To do this, we would separate the data into condition-specific segments and then estimate the MI for each segment. Following this, we calculate the difference between the MI estimated for each condition. However, each segment is effectively data sampled from a system conditioned on a third variable $\Theta$, i.e. $x,y \sim q(x,y | \theta)$, where $\theta \in \Theta$. As such, what we may initially call a MI $I(X;Y)$ in the condition $\theta$ is more formally a conditional MI, $I(X;Y|\Theta=\theta)$ -- when $\theta$ is one out of multiple possibilities for the system conditions. In comparison to an unconditioned MI, with samples taken across an ensemble of conditions, $\Theta$, this conditioning can remove \textit{redundancies} carried by both $X$ and $\Theta=\theta$ about $Y$, or add \textit{synergies} between $X$ and $\Theta=\theta$ about $Y$~\cite{williams2010,lizier2018}. We illustrate these effects in \fig{inside-support} and discuss them further in \secref{simulations}. This conditioning can potentially alter the meaning of the MI estimate for each condition, which can result in a misleading interpretation of the dependence between $X$ and $Y$ when comparing conditions. For example, conditions 1, 2 and 4 in \fig{state-switching} all have a similar MI, therefore the strength of dependence between $X$ and $Y$ appears to be the same in all of these conditions. However, conditions 2 and 4 have an offset (non-zero mean) that has not been captured. Relative to the probability space defined by conditions 1 and 3, which are centered on zero, conditions 2 and 4 have data points in low-probability regions, which would be informative in determining the statistical dependence.

In this work, we tackle the issues related to non-stationarity by introducing the \textit{cross} MI, described below.

\subsection{Cross MI and its properties}
We propose a new measure for the dependence between $X$ and $Y$ called the \textit{cross MI},
\begin{equation} \label{eq:cross-mi}
CI_{pq} = \mathbb{E}_{x,y \sim p(x,y)} \left\{ \log \left( \frac{q(x,y)}{q(x)q(y)} \right) \right\} = \mathbb{E}_{p} \left\{ i_q(x;y) \right\}.
\end{equation}
The cross MI evaluates the expected strength of dependence between $X$ and $Y$ exhibited in the data sampled from $p(x,y)$, referred to as the \textit{test distribution}. Cruicially, the strength of dependence for each sample taken from $p(x,y)$ is calculated using the probability distribution $q(x,y)$, referred to as \textit{reference distribution}. To contrast the two: $q(x,y)$ provides the reference or probabilistic model of how the variables are expected to relate in general, whilst $p(x,y)$ prescribes a specific set of samples for which we ask how this relationship is expressed. Similar to \eq{mi} and \eq{local-mi}, we use subscripts to refer to the probability distributions in the cross MI in \eq{cross-mi}.

Our definition of the cross MI (\eq{cross-mi}) is analogous to cross entropy \cite{bossomaier2016}, where `cross' refers to using a different probability distribution in calculating the measure from which the data is sampled. From a code length perspective, the cross entropy measures the average number of bits used for the test samples drawn from $p(x)$, assuming they were distributed as per the reference $q(x)$. Similarly, cross MI evaluates the code length saved for samples $x,y$ drawn from $p(x,y)$ assuming they were distributed as per the reference $q(x,y)$, as compared to assuming $x$ and $y$ were drawn independently from $q(x)$ and $q(y)$.

Like the conventional measures, cross MI is a sum and difference of cross entropies. When $p(x,y)=q(x,y)$ the cross MI is equivalent to the MI, i.e., $CI_{pp}=I_p$. Another property of \eq{cross-mi} is that if the reference distribution can be factorised as $q(x,y) = q(x) q(y)$, i.e., $X$ and $Y$ are independent in the reference, then the cross MI is zero by definition for any test distribution ($I_q = 0 \implies CI_{pq}=0$). Crucially, whilst $I_p \geq 0$ and $I_q \geq 0$, we can have $CI_{p,q} < 0$; this is because the pointwise MI for each sample $x,y \sim p(x,y)$ may be negative, and when taking the expectation over a different distribution to the reference the standard lower bound proof \cite[Sec. 2.6]{thomas2006} no longer applies. Additionally, the chain rule that applies to the conventional MI also holds for the cross MI (since it holds for each pointwise MI it averages over \cite{fano1961}). However, the data processing inequality does not hold (due to the possibility of the cross MI being negative).

If we adopt a Gaussian (linear) estimator for the MI, \appRef{linear-regression} presents specific mathematics on the form of the cross MI in this case. \appRef{linear-regression} goes on to demonstrate how the cross MI relates to cross validation in linear regression, specifically regarding how it improves on sum of squared residuals in measuring how well a model determined from the reference data performs on predicting the test data.

Note, the cross MI as defined here in \eq{cross-mi} is different to \cite{jeong2001, peralta2021} where the `cross' refers to calculating the MI across time and \cite{bugliarello2020} where a cross MI measure is defined as having different joint and marginal probability distributions in the logarithm (see \eq{mi}).

To calculate the cross MI, we need to specify the test data $x,y \sim p(x,y)$ and the reference distribution $q(x,y)$, which is estimated empirically from samples $x,y \sim q(x,y)$. For the test data, no distribution is explicitly estimated: if we have $N$ samples we would typically weight the local MI for each sample by $1/N$ in taking the expectation for $CI_{pq}$ in \eq{cross-mi}, in the same way that this is done for estimating $I_p$ (e.g., in the KSG estimator \cite{kraskov2004}).

\section{Simulations} \label{sec:simulations}
In this section we compare a conventional MI analysis to the cross MI using simulated data. We will consider different choices for the reference distribution and use a KSG estimator \cite{kraskov2004} in all calculations. All information-theoretic measures are expressed in units of nats.

\subsection{The cross MI can be used to handle non-stationary and online data}
Figure~\ref{fig:state-switching} illustrates the use of the cross MI in estimating the $X$-$Y$ dependence for a single online data point, which can be from a new (non-stationary) distribution. Here, it is not possible to calculate the MI for the test data point because we cannot accurately estimate the probability distribution it was sampled from. However, we are able to estimate a cross MI based on a particular choice for the reference distribution. Figure~\ref{fig:state-switching}B shows the MI for each system condition (i.e., of the reference data) and the cross MI of the test data point relative to the condition-specific reference distribution. The test data point is the same, however, the information it contains (local cross MI) is different depending on the reference distribution. When we use condition 1 for the reference distribution, this suggests the test data point has a high information content. However, considered relative to condition 2, the test data point is in a high-probability region for both joint and marginals so is not very informative. Condition 3 contains $X$ and $Y$ that are independent, which results in a small negative fluctuation away from zero in the cross MI estimate. Relative to condition 4, the test data point is atypical (with a low joint probability being highly surprising compared to the marginals) so has a negative cross MI. Relative to the full distribution (across all conditions, \fig{state-switching}C), we see the cross MI is not extreme enough to be statistically significant (see \appRef{statSigTesting}).

\subsection{Behaviour of the cross MI for different reference distributions that support the test data}
Next, we explore a scenario where we have multiple samples from the test distribution such that we are able to directly estimate the conventional MI of the test data. We will compare different choices for the reference distribution and see how this affects the cross MI. 

Figure~\ref{fig:inside-support} shows simulations where we have a linear relationship between $X$ and $Y$ in the test data and different relationships in the reference data. The test data can be viewed as being sampled from a conditioned state of the reference system, within a particular probability subspace. That is, $p(x,y) = q(x,y | \theta)$\footnote{Implicitly, the full distribution $q(x,y)$ is a accumulated over different conditions $q(x,y) = \sum_{\theta \in \Theta} q(\theta) q(x,y | \theta) = \sum_{\theta \in \Theta} q(\theta) p(x,y)$. The precise weighting $q(\theta)$ of different conditions plays a very important role in defining the reference distribution.}. In these simulations, the conditioning limits the $x$ range and variance of the test data.

\subsubsection{Synergy}
Figure~\ref{fig:inside-support}A shows the MI and cross MI for the test data when we have no $X$-$Y$ dependence in the reference data. We see the cross MI of test data relative to the reference data ($C_{pq}$) is much less than the MI of the test data ($I_p$). This is an example of a \textit{synergy}. Here, conditioning on a system condition $\theta$ provides a conditioned MI $I_p = I(X;Y|\theta)$, which reveals additional shared information between $X$ and $Y$ in the test data. Knowing that the data is sampled from a conditioned state adds additional information about $y$ than knowing $x$ alone (i.e., the variance in $y$ around $0.5x$ is more constrained than one would expect from the reference distribution alone). Measuring the MI relative to the reference distribution using the cross MI ($CI_{pq}$) results in a lower value than the MI for the test data ($CI_{pq} < I_p$) because we remove the synergy in the test data. A less severe synergy occurs in \fig{inside-support}B due to the reference encoding the same linear relationship here (albeit with more variance).

\subsubsection{Redundancy}
In \fig{inside-support}C we see the opposite, which is an example of a \textit{redundancy}. Here, conditioning on the system condition $\theta$ for $I_p = I(X;Y|\theta)$ has removed information shared between $x$, $y$ and $\theta$ in the test data. Knowing that the test data is sampled from a conditioned state already tells us a substantial portion of the information that $x$ carries about $y$ (i.e., the range of $y$ is now constrained in approximately $[0,1]$ rather than $[-2,2]$). This reduces the MI in the test data ($I_p$) compared to the reference data ($I_q$). However, once we measure the MI relative to the reference distribution using the cross MI ($CI_{pq}$), we observe that there is a strong $X$-$Y$ dependence in the test data. In comparison to $I_p$, this no longer excludes the redundant information from the condition $\theta$ and so is not reduced from $I_q$ across the full reference distribution. In fact, $CI_{pq}$ gives a higher value for the cross MI than the MI for the test data ($CI_{pq} > I_p$) because the specific $x$ values in the test distribution are more strongly related to the $y$ values than would be expected on average across $q(x,y)$.

\begin{figure}
    \centering
    \includegraphics[width=0.42\textwidth]{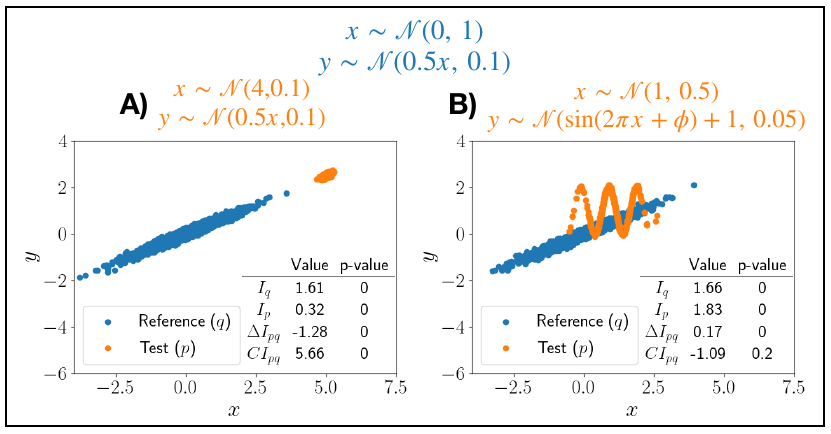}
    \caption{\textbf{Behaviour of the cross MI for different reference distributions that do not support the test data}. A) Test data (orange, 500 samples) outside of the range of the reference data (blue, 2000 samples). B) Different relationships between $X$ and $Y$ in the reference (blue, 2000 samples) and test (orange, 500 samples) data.}
    \label{fig:outside-support}
\end{figure}

\subsection{Behaviour of the cross MI for different reference distributions that do not support the test data}
Here, we consider scenarios where the reference distribution may not be an appropriate choice for the test data (referred to as being outside of the \textit{support} of the reference distribution). This can be because the test data is outside of the range of the reference data or because the form of the $X$-$Y$ dependence in the test data is not in the reference data.

Figure~\ref{fig:outside-support}A shows test data that extends the linear $X$-$Y$ dependence in the reference data to new $x$-values. This results in a very high value for the cross MI because the test data has a strong $X$-$Y$ dependence that is consistent with the reference data. However, the test data is beyond the $X$ and $Y$ range of the reference data, in low probability regions. Consequently, the cross MI becomes particularly sensitive to the estimator used. A linear Gaussian model estimator could still simply evaluate the reference Gaussian distribution in the range of the test data. In contrast, the model-free KSG \cite{kraskov2004} estimator used in this work adapts a box size used to estimate the reference probability density based on nearest neighbour distances from a given test data point. When the test data points are beyond the reference data (such as in \fig{outside-support}A), the estimated reference probability density for such model-free estimators becomes sensitive to the exact test data points used in the calculation. This can lead to a high variance in their cross MI estimates. The standard deviation of $C_{pq}$ for multiple samples of the test data is 0.26. In scenarios where the reference data cannot provide a reasonable model-free estimate for the probability density of the test data, the cross MI may not be appropriate with such estimators (see \secref{conclusions}).

In \fig{outside-support}B, we explore what happens if the dependency between $X$ and $Y$ in the test data is not present in the reference distribution. We simulate a nonlinear (sinusoidal) dependency for the test data and a linear dependency for the reference data. \fig{outside-support}B shows the cross MI is very negative, indicating the dependency in the test data is highly surprising (has a low probability) relative to the reference distribution. For the cross MI to pick up on an $X$-$Y$ dependence it must be shared in the reference and test data (see \secref{conclusions}).

\section{Conclusions} \label{sec:conclusions}
The cross MI provides a new measure for assessing the dependence between two random variables across different system conditions in a common reference probability space. It provides a measure for the strength of dependence in test data relative to a reference. In this section we discuss the choices that need to be made when calculating the cross MI and their consequences. We also discuss the limitations of this approach and some future applications.

\subsection{Choice of reference distribution}
One of the most important choices to be made is how to define the reference distribution. Ideally, the reference distribution will fully span and appropriately sample the accessible regions of the probability space of the system. This means the test data is sampled from a subspace in the reference distribution. The key questions are: does the reference distribution contain the dependencies between $X$ and $Y$ present in the test data? Does the reference distribution span the values of the test data? One way to assess whether this is the case is to compare the average nearest neighbour distance from the test data to the reference data to the average nearest neighbour distance within the reference data. If the test data is too far from the reference data, this may indicate that the reference distribution cannot adequately model it.

The test data being outside of the support of the reference data could potentially be resolved by including the test data in the estimation of the reference distribution. This ensures the reference distribution contains the $X$-$Y$ dependence in the test data and spans the test data. However, it comes at the cost of becoming sensitive to the ratio of test to reference data -- this is explored in \appRef{testRefDataRatio}.

\subsection{Conditional MI or cross MI}

Both the MI ($I_p$) and cross MI ($CI_{pq}$) provide a measure of the observed $X$-$Y$ dependence in the test data. The choice of which to use depends on the question the researcher wants to answer. Assuming the test distribution can be defined, then if one wishes to measure the $X$-$Y$ dependence \textit{with} the knowledge that the system is in a specific condition ($I_p$), then they accept that higher-order information from the conditioning is being included, and that any information redundant with the conditioning will be excluded. If they wish to measure the $X$-$Y$ relationship from a set of samples without reference to the system condition these have been sourced from ($CI_{pq}$), then one is accepting a purely pairwise evaluation of the relationship without taking higher-order information from the condition into account.
\subsection{Applications in task neuroimaging data analysis}
The ability to compare the the $X$-$Y$ dependence in different system conditions is particularly useful in the analysis of neuroimaging data from task experiments, where we record data from participants whose brains may transition between different task conditions. Here, we briefly discuss how the cross MI can be used to study different task conditions.

The interaction between two brain regions $X$ and $Y$ will change as a function of the overall brain state, perhaps being the cognitive task the subject is performing. In considering the interaction of a pair of regions on their own during one particular task, the cross MI ($CI_{pq}$) may be used to evaluates that pair's activity with reference to the ensemble of dynamics they experience. This is a measurement of their pairwise interaction that contains no other explicit encoding of the task or state beyond their own dynamics. The MI ($I_p$) on the other hand takes the task or state as a given for the observer. This could be taken to represent an explicit encoding of the task provided by some third party region gating the interaction; conditioning on such a signal, takes this beyond a pairwise measure of the interaction.

The key choice to be made is the data for the reference distribution. Ideally, were recordings available for long-term periods (i.e. weeks, months) one could use these to build a reference model for the ensemble of brain dynamics across many tasks encountered ``in the wild'' according to their natural relative likelihoods. More pragmatically, one might use a full task recording, which since task design usually alternates between different conditions and rest periods ensuring the reference distribution contains all conditions of interest. 

\subsection{Extension to other information-theoretic measures}
The cross MI can be trivially extended to cross conditional mutual information $CI_{pq}(X;Y|Z)$, and any other information-theoretic measures based on these. These include constructing a ``cross'' active information storage \cite{lizier2012},
$
CAIS_{pq} = CI_{pq}(X_t; X_{<t}),
$ 
as well as ``cross'' transfer entropy \cite{schreiber2000,bossomaier2016},
$
CTE_{pq} = CI_{pq}(X_{<t}; Y_t | Y_{<t}).
$

\section*{Code Availability}
The MI and cross MI were computed using JIDT~\cite{lizier2014}. Code to reproduce the results presented can be found here: \url{https://github.com/InfoDynamicsTeam/CrossMI}.

\section*{Funding Acknowledgments}
CG, JMS, BDF and JTL were supported through the Australian Research Council Discovery Project grant no.
DP240101295.

\ifthenelse{\not\boolean{proceedings}}{

\begin{figure*}
    \centering
    \includegraphics[width=\textwidth]{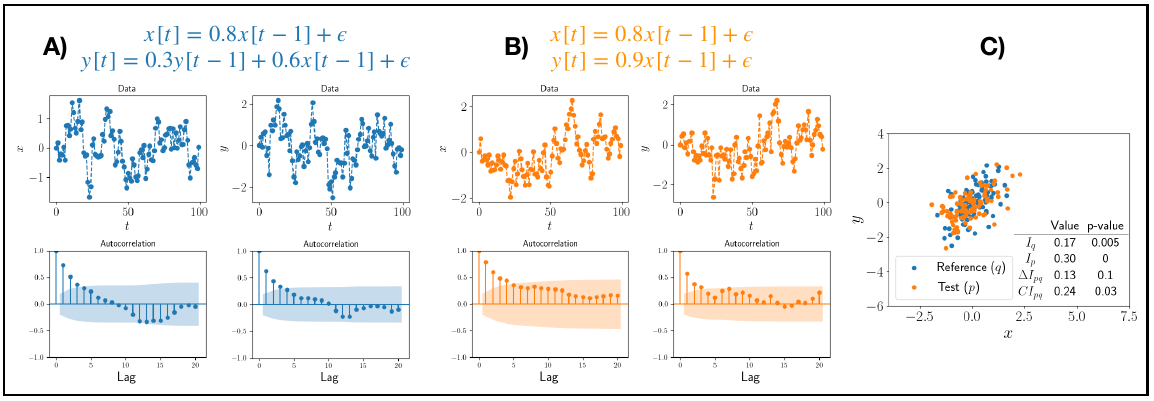}
    \caption{\textbf{The proposed statistical significance testing can be applied to autocorrelated data.} Simulated data (100 samples) and autocorrelation function for the reference (A) and test data (B). Information theory measures calculated with the data (C). A block length of 5 samples was used to perform statistical significance testing.}
    \label{fig:autocorr}
\end{figure*}

\begin{figure*}
    \centering
    \includegraphics[width=\textwidth]{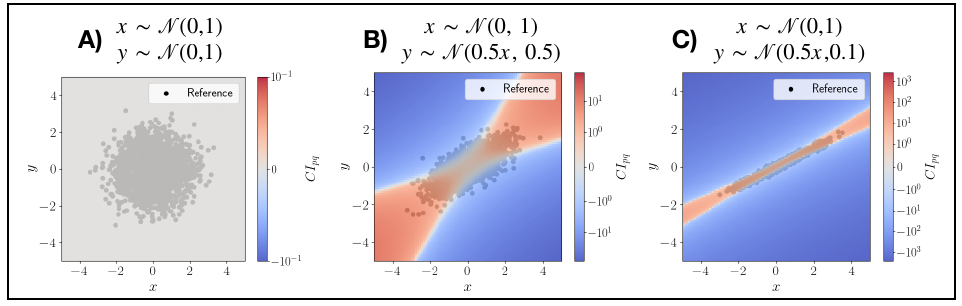}
    \caption{\textbf{Heatmaps for local cross MI calculated analytically assuming a normal reference distribution}. Difference choices for the reference data (grey, 2000 samples) are shown in the background in A, B and C, with the distributions they are drawn from displayed above the plots.}
    \label{fig:analytical-cross-mi}
\end{figure*}

\begin{figure*}
    \centering
    \includegraphics[width=\textwidth]{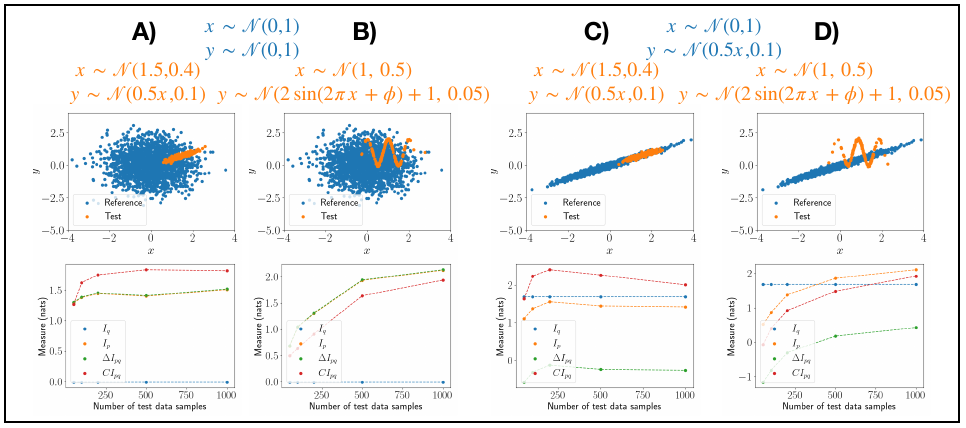}
    \caption{\textbf{The cross MI scales with number of samples in the test data if we include the test data in estimating the reference distribution.} Top: illustrative data simulated for each system. Bottom: information theoretic measures as a function of number of test samples. All examples simulated 2000 samples from the reference system. Here, we are using both the reference (blue) and test (orange) data to estimate the reference distribution $q(x,y)$.}
    \label{fig:sample-scaling}
\end{figure*}

\appendices

\section{Statistical Significance Testing}
\label{app:statSigTesting}
We use statistical significance testing to determine whether the value of the measure we observed could be due to chance. For the standard MI, this means asking whether the statistic $I_p$ is statistically significant against surrogate measurements that we would obtain from null models where the marginal distributions $p(x)$ and $p(y)$ are retained whilst the joint distribution relating the variables $p(x,y)$ is not.

In moving to consider the cross MI, there are a number of different ways we can ask whether our measures are statistically significant compared to a null model. This section presents what those questions are and how we address them.

For all of these questions, we use non-parametric permutation tests. Here, we build a null distribution by permuting the original data. Looking up where the observe statistic lies in the null distribution indicates how likely the statistic is under the assumption that the null hypothesis is true. A problem often encountered in performing statistical significance testing with time series data is autocorrelation of the samples, such that they are not independent (e.g., see \cite{cliff2021}). This aspect of the data must be preserved in the permutations used to build the null distribution to ensure the variance of the null distribution is correctly estimated. We adopt a \textit{block shuffle} permutation to build the null distribution. This is a well known approach for performing permutation tests with time series data~\cite{lahiri2013}. The block shuffle permutation separates a time series into continuous blocks of equal size and shuffles the position of the blocks to produce a surrogate time series.

The procedures described below works well provided the block length is greater than the typical autocorrelation length in the test data and if there are enough permutations to obtain a good resolution for the null distribution. The block length (number of samples in each block) is chosen to be greater than the typical autocorrelation length in the data ($x$ or $y$, whichever is larger). The data simulated in this work did not contain autocorrelation, so we used a block length of one. In \fig{autocorr}, we perform statistical significance testing on autocorrelated data. We used 200 permutations to build the null distribution.

\subsubsection*{MI: is $|I|$ non-zero?}
We are interested in testing whether an estimate for the MI ($I$) is significantly different from zero by breaking the dependence between $X$ and $Y$ while preserving any other characteristics, such as autocorrelation, in the data. This is done using the following steps:
\begin{enumerate}
\item Block shuffle the $x$ time series to break the dependence between $X$ and $Y$. This provides a surrogate $x$ time series.
\item Calculate MI using the surrogate $x$ time series and the non-permuted $y$ time series.
\item Repeat a number of times to build a null distribution for $I=0$.
\item Look up where the MI for the non-permuted data lies in the null distribution. This provides a $p$-value for $|I| > 0$.
\end{enumerate}

\subsubsection*{Differences in MI: is $\Delta I_{pq} = |I_p - I_q|$ non-zero?}
We are interested in testing whether the observed difference in MI is significantly different from zero. We do this by randomly flipping the system label, referred to as a `sign-flip' permutation. The idea is if the difference is zero, multiplying by -1 should not affect the statistic. This is done using the following steps:
\begin{enumerate}
\item Calculate the local MI for each system condition of interest: $i_q$ and $i_p$.
\item Separate the local MI into blocks and average the values in each block: $i^b_q$ and $i^b_p$. There may be a different number of blocks for the reference and test system.
\item Randomly assign the system label to each block MI with a probability reflecting the fraction of blocks from each system.
\item Take the difference in the mean across blocks: $\Delta I_{pq} = \left< \tilde{i}^b_p \right> - \left< \tilde{i}^b_q \right>$, where $\tilde{i}^b_q$ and $\tilde{i}^b_p$ are the shuffled block MIs and $\left< . \right>$ denotes the average.
\item Repeat to build a null distribution for $\Delta I_{pq} = 0$.
\item Look up where the non-permuted difference in MI ($\Delta I_{pq}$) lies in the null distribution. This provides a $p$-value for $|\Delta I_{pq}| > 0$.
\end{enumerate}

\subsubsection*{Cross MI: is $|CI_{pq}|$ non-zero?}
We are interested in whether an observed value for the cross MI is significantly different from zero, indicating there is a dependence between $X$ and $Y$ in the test data relative to the reference distribution. We can do this by either breaking the dependence between $X$ and $Y$ in the test data or the reference data. Generally, we are interested in the $X$-$Y$ dependence in the test data and want to break the dependence in the test data. However, in scenarios where there's limited test data (e.g., with online data), we break the dependence in the reference data instead. We use the following steps:
\begin{enumerate}
\item Split the $x$ time series in the test (or reference) data into blocks and shuffle to obtain a surrogate $x$ test time series.
\item Calculate the surrogate cross MI using the surrogate $x$ test time series and non-permuted test $y$ time series along with the non-permuted reference data. Note, we also rely on the dependence between $X$ and $Y$ in the reference data being non-zero to obtain a non-zero cross MI - see \secref{theory}.
\item Repeat to build a null distribution.
\item Look up what percentile the non-permuted cross MI value falls in the null distribution to obtain a $p$-value.
\end{enumerate}

\section{Relation to linear regression}
\label{app:linear-regression}
A common model for the relationship between two variables $X$ and $Y$ is linear regression \cite{bishop2006}:
\begin{equation} \label{eq:linear-regression}
y = \beta x + \gamma + \epsilon,
\end{equation}
where $\beta = \frac{\sigma_{XY}^2}{\sigma_X^2} = \rho_{XY} \frac{\sigma_{Y}}{\sigma_X}$ (for variances $\sigma_X$ and $\sigma_Y$, covariance $\sigma_{XY}^2$ and correlation $\rho_{XY}$) is a constant known as a \textit{regression coefficient}, $\gamma$ is a constant, and $\epsilon$ is a normally distributed \textit{residual}. Modelling the reference data with \eq{linear-regression} is equivalent to estimating a posterior model for the conditional probability distribution,
\begin{equation}
q(y | x) = \mathcal{N}(\mu_{Y|x}, \sigma^2_{Y|X}),
\end{equation}
where $\mu_{Y|x} = \beta x + \gamma$ is the mean for variable $Y$ given a realisation $X=x$, and $\epsilon \sim \mathcal{N}(0, \sigma^2_{Y|X})$ (which is independent of $X=x$). Note that $\sigma^2_{Y|X} = \sigma^2_{Y}(1 - \rho_{XY}^2)$. This conditional distribution can be compared to a prior model for the marginal distribution of $Y$ from the reference data, $q(y) = \mathcal{N}(\mu_Y, \sigma^2_{Y})$, in order to calculate the various MI quantities, including cross MI for new test data, analytically\footnote{Here, we are using a `Gaussian estimator' for the cross MI.}. Substituting $q(x,y) = q(y|x) q(x)$ into \eq{cross-mi}, we have the other standard form for the pointwise MI for a sample $(x,y)$:
\begin{equation} \label{eq:conditional-pointwise-mi}
i_{q}(x,y) = \log \left( \frac{q(y|x)}{q(y)} \right),
\end{equation}
and substituting normal distributions for the reference,
\begin{equation}
    \begin{aligned}
    q(y | x) &= \frac{1}{\sqrt{2\pi \sigma^2_{Y|X}}} \exp \left( -\frac{1}{2 \sigma^2_{Y|X}} ( y - \mu_{Y|x} )^2 \right), \\
    q(y) &= \frac{1}{\sqrt{2\pi \sigma^2_{Y}}} \exp \left( -\frac{1}{2 \sigma^2_{Y}} ( y - \mu_{Y} )^2 \right), \\
    \end{aligned}
\end{equation}
into \eq{conditional-pointwise-mi}, we get
\begin{equation} \label{eq:analytical-pointwise-mi}
\begin{aligned}
i_{q}(x,y) = &-\frac{1}{2} \log \left( \frac{\sigma^2_{Y|X}}{\sigma^2_Y} \right) \\
 &+ \frac{1}{2}  \left\{ \frac{(y-\mu_Y)^2}{\sigma^2_Y} - \frac{(y-\mu_{Y|x})^2}{\sigma^2_{Y|X}} \right\}.
\end{aligned}
\end{equation}
The first term in \eq{analytical-pointwise-mi} is the expected MI of the reference data $I_q(X;Y)$, simplifying to the well-known form $-\frac{1}{2}\log(1-\rho_{XY}^2)$ \cite[sec. 8.5]{thomas2006} (noting as above $\frac{\sigma^2_{Y|X}}{\sigma^2_{Y}} = 1-\rho_{XY}^2$, which is also referred to as the ``fraction of variance unexplained'').
The second term adds a correction to the underlying average MI, based on comparing the squared relative residuals from the prior and posterior models. In each case, the residual is the difference between the actual $y$ and the prediction from the model of the reference data ($\mu_Y$ for prior model and $\mu_{Y|x}$ for posterior), which is then normalised by the expected variance ($\sigma_Y$ for the prior and $\sigma_{Y|X}$ for the posterior) to make it a relative term.
The correction term can be seen to increase (decrease) the MI by the amount by which the relative error of the posterior model was smaller (larger) than that of the prior model, for this sample $(x,y)$.
Of course, when the average is taken over $q$ the numerators cancel the denominators of each correction term (by definition), and we are left with the first term only for the average MI.

The cross MI when using the Gaussian model is then the average of these pointwise MI values from \eq{analytical-pointwise-mi} over $p$:
\begin{equation}\label{eq:analytical-cross-mi}
\begin{aligned}
CI_{pq} = &-\frac{1}{2} \log \left( \frac{\sigma^2_{Y|X}}{\sigma^2_Y} \right) \\
 &+ \frac{1}{2} \mathbb{E}_{x,y \sim p(x,y)} \left\{ \frac{(y-\mu_Y)^2}{\sigma^2_Y} - \frac{(y-\mu_{Y|x})^2}{\sigma^2_{Y|X}} \right\}.
\end{aligned}
\end{equation}
This of course contains the MI of the reference data as the first term $I_q(X;Y)$, with the subsequent term averaging the aforementioned corrections (based on the residuals of each model) over all samples in $p$.

Conventionally, \textit{cross validation} is used to quantify how well a linear regression model determined from the reference (or training) data fits new out-of-sample test data using the residuals $\epsilon_i$ for each test sample $(x_i,y_i)$, $\epsilon_i = y_i- \mu_{Y|x_i}$. Specifically, the sum of squared residuals $\sum_i \epsilon_i^2 = \sum_i (y_i- \mu_{Y|x_i})^2$ is computed as a measure of the discrepancy between model predictions and actual test data, with a smaller value indicating a better fit. Interestingly, the cross MI contains this term in \eq{analytical-cross-mi}, normalised to $\sigma^2_{Y|X}$, and with an opposite sign, so smaller residuals serve to increase $CI_{pq}$ indicating stronger applicability of the model determined from the reference data to the test data.
This aligns with our earlier interpretations of the cross MI as measuring how strongly the relationship defined in the reference data is expressed in the test data, evaluated here with a linear regression model.
This raises the question of how the properties of cross MI (with the normalisation of the sum of squared residuals plus extra terms) may compare to the sum of squared residuals alone as a measure of model fit to the test data.
Recalling that:
\begin{enumerate}
\item the pointwise MI in \eq{conditional-pointwise-mi} is derived as the unique form to measure the information a sample $x$ provides about $y$ given the reference distributions $q(y)$ and $q(y|x)$ fulfilling certain axioms (including a chain rule) \cite{fano1961}, and
\item that such information is a measure of the quality of prediction in terms of how the sample $x$ constrains our expectation of the probabilities for $y$ given $x$ (see e.g. \cite{finn2018}),
\end{enumerate}
then one would conclude that the cross MI (in averaging the pointwise MI over the test data) would be the unique measure of quality of prediction that would satisfy these same axioms.
This includes, notably, offering a chain rule over multiple predictors $X_1,X_2,\ldots$ for $Y$ to provide consistency in how the quality of prediction accumulates across them.
The normalisation of the sum of squared residuals in the cross MI, as well as the extra terms, appears necessary then in order to satisfy such axioms.

To experimental results applying the linear model then, \fig{analytical-cross-mi} shows the pointwise cross MI calculated using the linear model in \eq{analytical-cross-mi} for the reference distributions used in \fig{inside-support}. \fig{analytical-cross-mi}A, where $X$ and $Y$ are independent in the reference distribution, shows the cross MI is zero everywhere, because  so knowing $x$ provides no information regarding $y$ anywhere. This is consistent with our earlier statement that $I_q = 0 \implies CI_{pq}=0$. \fig{analytical-cross-mi}B and C show the local cross MI when there is a relationship in the reference data. Note that there are two gradients for the pointwise cross MIs here:
\begin{enumerate}
\item The MI increases as we approach the trendline $y = \beta x + \gamma$ for the posterior model, since $X$ becomes more strongly predictive of $Y$ using the model along this gradient. Quantitatively, for fixed $y$ (constant $q(y)$), moving towards the trendline means $q(y|x)$ increases which directly increases $i_q(x,y)$ in \eq{analytical-pointwise-mi}.
\item The MI increases along or parallel to the trendline for the posterior model, when moving away from the perpendicular bisector through the variables' means. Quantitatively, $q(y|x)$ is fixed along these parallels to the trendline, whilst $q(y)$ decreases along these perpediculars; this directly increases $i_q(x,y)$ in \eq{analytical-pointwise-mi}.
\end{enumerate}

Finally, we see that assuming a normal distribution (i.e., using a `Gaussian estimator') is consistent with the results from the KSG estimator used in \fig{inside-support} for various test distributions for the same reference distributions.
For example, we see that the test samples in \fig{inside-support}B and C sit inside the positive pointwise MI areas of the heatmaps in \fig{analytical-cross-mi}B and C respectively, leading to positive cross MIs there.
In contrast, we see that the test samples in \fig{outside-support}B primarily sit in negative pointwise MI areas of the heatmap in \fig{analytical-cross-mi}C (which has the corresponding reference distribution), leading to a negative cross MI there.

\section{Sensitivity to the test/reference data ratio}
\label{app:testRefDataRatio}
\fig{sample-scaling} shows how different information-theoretic measures change with the number of test samples for various of scenarios when we use both the test and reference data to estimate $q(x,y)$. Generally, the test MI ($I_p$) and the cross MI ($CI_{pq}$) increases with the number of test samples. The sensitivity in $CI_{pq}$ arises due to the test data being included in the reference distribution. The change in $I_p$ and $CI_{pq}$ depends on the nature of the dependency between $X$ and $Y$ in the test system and its relation to reference system. This may not be an issue. For example, this would be fine if we are simply interested in calculating the cross MI of the test system relative to the reference ($CI_{pq}$). This is more problematic if we are trying to compare the cross MI for two different systems. If we include the test data in the reference distribution, we may find differences in the cross MI that are simply due to having a different number of samples in each test dataset. Care must be taken to ensure each system is represented in defining the reference distribution when comparing different systems.

}{}

\end{document}